\documentclass[fleqn,usenatbib]{mnras}

\usepackage{newtxtext,newtxmath}

\usepackage[T1]{fontenc}
\usepackage{ae,aecompl}

\usepackage{graphicx}	
\usepackage{amsmath}	
\usepackage{amssymb}	

\title[GW background from TDEs]{The gravitational wave background signal from tidal disruption events}

\author[M. Toscani et al.]{
Martina Toscani,$^{1}$\thanks{E-mail: martina.toscani@unimi.it}
Elena M. Rossi$^{2}$,
Giuseppe Lodato,$^{1}$
\\
$^{1}$Dipartimento di Fisica, Universit\`a Degli Studi di Milano, Via Celoria, 16, Milano, 20133, Italy\\
$^{2}$Leiden Observatory, Leiden University, PO Box 9513, 2300 RA, Leiden, the Netherlands
}

\date{Accepted XXX. Received YYY; in original form ZZZ}

\pubyear{2020}

\date{Accepted XXX. Received YYY; in original form ZZZ}

\pubyear{2020}

\begin{document}
\label{firstpage}
\pagerange{\pageref{firstpage}--\pageref{lastpage}}
\maketitle

\begin{abstract}
\textcolor{black}{In this paper we derive the gravitational wave stochastic background from tidal disruption events (TDEs). We focus on both the signal emitted by main sequence stars disrupted by super-massive black holes (SMBHs) in galaxy nuclei, and on that from disruptions of white dwarfs by intermediate mass black holes (IMBHs) located in globular clusters. We show that the characteristic strain $h_{\rm c}$'s dependence on frequency is shaped by the pericenter distribution of events within the tidal radius, and under standard assumptions $h_{\rm c} \propto f^{-1/2}$. This is because the TDE signal is a burst of gravitational waves at the orbital frequency of the closest approach. In addition, we compare the background characteristic strains with the sensitivity curves of the upcoming generation of space-based gravitational wave interferometers: the Laser Interferometer Space Antenna (LISA), TianQin, ALIA, the DECI-hertz inteferometer Gravitational wave Observatory (DECIGO) and the Big Bang Observer (BBO). We find that the background produced by main sequence stars might be just detected by BBO in its lowest frequency coverage, but it is too weak for all the other instruments. On the other hand, the background signal from TDEs with white dwarfs will be within reach of ALIA, and especially of DECIGO and BBO, while it is below the LISA and TianQin sensitive curves. This background signal detection will not only provide evidence for the existence of IMBHs up to redshift $z\sim 3$, but it will also inform us on the number of globular clusters per galaxy and on the occupation fraction of IMBHs in these environments.}

\end{abstract}



\vspace{2 cm}
\section{Introduction}
Tidal disruption events (TDEs) are transient astronomical events that occur when a star, wandering too close to a black hole (BH), gets disrupted by the tidal forces due to the hole, that overwhelm the stellar self-gravity (see \citealt{Rees:88aa} and \citealt{Phinney:89aa}). After the phase of disruption, about half of the star circularizes around the hole and is expected to form an accretion disc (\textcolor{black}{\citealt{Hayasaki:13aa}}, \citealt{Shiokawa:15aa}, \citealt{Bonnerot:16aa} and \citealt{Hayasaki:16aa}), while the other half escapes on hyperbolic orbits with different energies. These phenomena are very luminous electromagnetic sources (see, e.g., \citealt{Komossa:08aa}, \citealt{Bloom:11aa}, \citealt{Komossa:15aa}, \citealt{Gezari:17aa}), with a luminosity decay that, at late times in soft-X rays, might be expected to decline as $t^{-5/3}$ (\citealt{Lodato:09aa}, \citealt{Lodato:11aa}, \citealt{Guillochon:13aa}).\\
\indent During these events, we do not only expect electromagnetic emission, but also gravitational wave (GW, \citealt{Einstein:18aa}) production. In particular, three different processes emit GWs during TDEs. First, there are GWs generated by the time-varying mass quadrupole of the star-BH system. This emission has been investigated initially by \citet{Kobayashi_2004}. They study the tidal disruption of a Sun-like star by a super-massive black hole (SMBH) with $\text{M}_{\bullet}\approx 10^6\text{M}_{\sun}$, obtaining a GW strain $h \approx 10^{-22}$ if the BH is not-rotating, while $h\approx 10^{-21}$ if the SMBH is spinning. Similarly,  \citet{Rosswog:09ab}, \citet{Haas:12aa} and \citet{Anninos:18aa} explore ultra-close TDEs of white dwarfs (WDs) by intermediate mass black holes (IMBHs), that might have been observed \citep{Lin:18aa,Peng:19aa,Lin:20aa}. A WD with mass $\approx 1\text{M}_{\sun}$ and radius $\approx 10^{9}\,\text{cm}$ is expected to generate a strain $h\approx 10^{-20}$, if disrupted by a $10^3\,\text{M}_{\sun}$ IMBH at $\approx 20\,\text{kpc}$ from us. Secondly, there are GWs produced by the variation of the internal quadrupole moment of the star as it gets compressed and stretched by the tidal forces when passing through the pericenter. In particular, \citet{Guillochon:09aa} study this emission for a Sun-like star disrupted by a $10^6\,\text{M}_{\sun}$ SMBH numerically, while \citet{Stone:13aa} focus more on the analytical investigation of this emission both from main sequence (MS) stars and WDs tidally disrupted. They both show that all being equal, this signal is in general one order of magnitude lower than that produced by the star-SMBH system. The two signals become comparable only if the TDEs are highly penetrating. Lastly, emission of GWs may arise after the circularization of debris around the BH. GWs may be produced by an unstable accretion disc where the Papaloizou-Pringle instability occurs (see \citealt{Papaloizou:84aa}, \citealt{Blaes:86aa})
. This is a global, non axi-symmetric, hydrodynamical instability that generates a localized overdensity that orbits the BH and gradually spreads out. This clump is the source of GWs (\citealt{Vanputten:01aa}, \citealt{Kiuchi:11aa}, \citealt{Toscani:19aa}, \citealt{Vanputten:19aa}). In particular, \citet{Toscani:19aa} show that for a $1\,\text{M}_{\sun}$ torus around a $10^6\,\text{M}_{\sun}$ non-rotating SMBH, this signal is around $10^{-24}$, with a frequency $\approx\,$mHz.\\
\indent All these studies focus on the detection of GW emission from single disruption events and they all show that, although these signals are in the Laser Interferometer Space Antenna (LISA, \citealt{Amaro-Seoane:17aa}) frequency band, they are quite weak, so it will be unlikely for LISA to detect them. In this paper we explore the GW signal produced by the entire cosmic population of TDEs (signal from the BH-star system), that would result in a stochastic background. We will investigate both the background associated with TDEs of MS stars with SMBHs and the one generated by the disruption of WDs by IMBHs. We then compare these signals with the sensitivity curves of the next generation of GW interferometers, i.e. LISA, TianQin (\citealt{Luo:16aa}), ALIA (\citealt{Bender:13aa}, \citealt{Baker:19aa}), the DECI-hertz inteferometer Gravitational wave Observatory (DECIGO, \citealt{Sato:17aa}) and the Big Bang Observer (BBO, \citealt{Harry:06aa}). The detection and characterisation of this background signal would provide unique information on TDE rates and on the hidden SMBH quiescent population, including that of the elusive IMBHs up to redshift $z \sim 3$. \\
\indent The structure of this paper is the following: in section \ref{sec:2} we describe the basic theory of TDEs and the derivation of the GW background in the most general case; in section \ref{sec:3} we describe in detail our method; in section \ref{sec:4} we illustrate our results, while in section \ref{sec:5} and section \ref{sec:6} we discuss the work done and we draw our conclusions.

\section{Theory}
\label{sec:2}

\subsection{Gravitational signal from tidal disruption events}
\label{par:TDE_theory}
Let us consider a star of mass $M_{*}$ and radius $R_{*}$, on a parabolic orbit around a non-spinning black hole of mass $M_{\rm h}$. The TDE takes place when the tidal forces due to the BH overcome the stellar self-gravity. For the purpose of the present paper, it is sufficient to use the so-called \textit{impulse approximation}, which means that the star interacts with the hole only at the pericenter $r_{\rm p}$, where it gets disrupted. Because of the varying quadrupole moment of the BH-star system, we expect a GW burst at pericenter \citep{Kobayashi_2004}. A simple estimate of the (maximum) GW strain emitted by the source is (see, e.g., \citealt{Thorne:98aa})
\begin{align}
  h \approx \frac{1}{\textcolor{black}{d}}\frac{4G}{c^2}\frac{E_{\rm kin}}{c^2},
\label{eq:ekin}
\end{align}
where $\textcolor{black}{d}$ is the \textcolor{black}{distance} of the source \textcolor{black}{from Earth} and $E_{\rm kin}$ is the kinetic energy of the star, since due to the high mass ratio between the BH and the star we can consider the BH at rest in the centre of mass frame (see appendix A of \citealt{Toscani:19aa} for a more detailed discussion on this assumption). Assuming the star to be a point-like particle in Keplerian motion, we write $E_{\rm kin}$ as
\begin{align}
    E_{\rm kin}=\textcolor{black}{M_*\frac{GM_{\rm h}}{r_{\rm p}}}.
\end{align}
Thus, the GW strain becomes
\begin{align}
    h&\approx \beta\times\frac{\textcolor{black}{r_{\rm s}}r_{\rm s*}}{r_{\rm t}\textcolor{black}{d}}\nonumber\\
    &\approx \beta \times \textcolor{black}{2}\times 10^{-22}\left(\frac{M_{*}}{\text{M}_{\sun}} \right)^{4/3}\left(\frac{M_{\rm h}}{10^6\text{M}_{\sun}} \right)^{2/3}\left(\frac{R_{*}}{\text{R}_{\sun}} \right)^{-1}\left(\frac{\textcolor{black}{d}}{16\text{Mpc}} \right)^{-1},\label{eq:burst}
\end{align}
with an associated frequency 
\begin{align}
    f&\approx\frac{\beta^{3/2}}{2{\rm \pi}}\left(\frac{GM_{\rm h}}{r^3_{\rm t}}\right)^{1/2}\nonumber\\
    &\approx \beta^{3/2}\times 10^{-4}\,\text{Hz}\times\left(\frac{M_{*}}{\text{M}_{\sun}} \right)^{\textcolor{black}{1/2}}\left(\frac{R_{*}}{\text{R}_{\sun}} \right)^{\textcolor{black}{-3/2}},\label{eq:freqkepl}
\end{align}
where we have introduced the \textcolor{black}{Schwarzschild radius $r_{\rm s}$ of the BH}, the Schwarzschild radius $r_{\rm s*}$ of the star, and the maximum pericenter distance for tidal disruption (a.k.a {\em tidal radius}) 
\begin{align}
    r_{\rm t}&\approx R_{*}\left(\frac{M_{\rm h}}{M_{*}} \right)^{1/3}\\
    & \approx 7\times 10^{12}\,\text{cm}\left(\frac{R_{*}}{\text{R}_{\sun}}\right)^{\textcolor{black}{+1}}\left(\frac{M_{*}}{\text{M}_{\sun}}\right)^{-1/3}\left(\frac{M_{\rm h}}{10^{6}\text{M}_{\sun}}\right)^{1/3}.
\end{align}
In fact, $r_{\rm t}$ should have also a numerical factor of a few, due to the internal structure of the star, relativistic effects in the process of disruption and other physical details of the system. We take this factor to be 1 for simplicity. The penetration factor $\beta$ is defined as
\begin{align}
    \beta\doteq\frac{r_{\rm t}}{r_{\rm p}}.
\end{align}
This factor varies between a minimum value $\beta_{\rm min}=1$ (i.e. $r_{\rm p}=r_{\rm t}$), and a maximum value $\beta_{\rm max}=r_{\rm t}/r_{\rm s}$, when the pericenter is equal to the BH Schwarzschild radius. Within this radius the star is directly swallowed rather than disrupted by the BH. For $\beta=1$ and a Sun-like star disrupted by a $10^6\,\text{M}_{\sun}$ static BH at $\approx 16\,\text{Mpc}$ from us, equations \eqref{eq:burst}-\eqref{eq:freqkepl} give $h\approx 10^{-22}$ and $f\approx 10^{-4}\,\text{Hz}$ (cf. \citealt{Kobayashi_2004}).

\subsubsection{White dwarfs}
In the rest of this paper, we assume that the WD mass is fixed and equal to $M_*=0.5\text{M}_{\sun}$. Following \citet{Shapiro:83aa}, a WD with this mass has a radius $R\approx 10^{-2}R_{\sun}$. The upper limit on the mass of the BH involved in the disruption is found by $r_{\rm t}>r_{\rm s}$ to be
\begin{align}
M_{\rm h}\lesssim 2\times 10^5\text{M}_{\sun}.
\end{align}
Thus, we can take $10^3\text{M}_{\sun}\leq M_{\rm h} \leq 10^5\text{M}_{\sun}$ as a reasonable range for the IMBH mass. Events with smaller BH masses will emit signals at least 100 times dimmer (see equation \ref{eq:burst}) and therefore we ignore them. We assume that these IMBHs reside in GCs.\\
\indent Considering what said above for the $\beta$ parameter, we obtain
\begin{align}
    &1\leq\beta\lesssim 29\,\,\text{if}\,\,M_{\rm h}=10^3\text{M}_{\sun},\nonumber \\
    &1\leq\beta\lesssim 6\,\,\,\,\,\text{if}\,\,M_{\rm h}=10^4\text{M}_{\sun},\\
    &1\leq\beta\lesssim 1.4\,\text{if}\,\,M_{\rm h}=10^5\text{M}_{\sun},\nonumber
\end{align}
and, as a result, we have the following limits on the GW strain and frequency, assuming an average distance of $16\,\text{Mpc}$ (Virgo Cluster)
\begin{align}
   \textcolor{black}{8}\times 10^{-23}\lesssim\,  &h_{3} \lesssim \textcolor{black}{2.4}\times 10^{-21}, \,\, 7\times 10^{-2}\text{Hz}\lesssim f_{3} \lesssim 11\text{Hz}\nonumber ,\\
    \textcolor{black}{4}\times 10^{-22}\lesssim \, &h_{4} \lesssim \textcolor{black}{2.2}\times 10^{-21}, \,\, 7\times 10^{-2}\text{Hz}\lesssim f_{4} \lesssim 1\text{Hz},\\
     \textcolor{black}{2}\times 10^{\textcolor{black}{-21}}\lesssim \, &h_{5} \lesssim \textcolor{black}{2.2}\times 10^{-21}, \,\, 7\times 10^{-2}\text{Hz}\lesssim f_{5} \lesssim 0.1\text{Hz}\,\,\nonumber,
\end{align}
where the index $h_{\rm x}$ ($f_{\rm x}$) means that we consider $M_{\rm h}=10^{\rm x}\text{M}_{\sun}$.

\subsubsection{Main sequence stars}
For main sequence (MS) stars, we assume $1\,\text{M}_{\sun}\leq M_{*} \leq \,100\,\text{M}_{\sun}$. Considering the scaling relation $M_{*}\approx R_*$ and a star with $1\,\text{M}_{\sun}$ and $1\,\text{R}_{\sun}$ we get
\begin{align}
M_{\rm \bullet}\lesssim 10^8\text{M}_{\sun}.
\end{align}
Thus, we take $10^6\,\text{M}_{\sun}\leq M_{\bullet} \leq \,10^8\,\text{M}_{\sun}$ as the BH mass range (note that we use $M_{\rm h}$ to refer to the mass of IMBHs and $M_{\bullet}$ to refer to SMBHs). Since these BHs are super-massive, we expect them to reside in galactic nuclei. For a Sun-like star we get the following intervals for $\beta$
\begin{align}
    &1\leq\beta\lesssim \textcolor{black}{23}\,\,\text{if}\,\,M_{\bullet}=10^6\text{M}_{\sun},\nonumber \\
    &1\leq\beta\lesssim \textcolor{black}{5}\,\,\,\,\,\text{if}\,\,M_{\bullet}=10^7\text{M}_{\sun},\\
    &\beta \approx 1\,\,\,\,\,\,\,\,\,\,\,\,\,\,\,\,\,\text{if}\,\,M_{\bullet}=10^8\text{M}_{\sun},\nonumber
\end{align}
and the strain for a source at 16 Mpc and its frequency span in the following ranges
\begin{align}
   &\textcolor{black}{2}\times 10^{-22}\lesssim\,  h_{6} \lesssim \textcolor{black}{5}\times 10^{-21}, \,\,\,\,\,\,\,\,\,\,\,\, 10^{-4}\text{Hz}\lesssim f_{6} \lesssim \textcolor{black}{ 10^{-2}}\text{Hz}\nonumber ,\\
    &\textcolor{black}{9}\times 10^{-22}\lesssim \, h_{7} \lesssim \textcolor{black}{4}\times 10^{-21}, \,\, \,\,\,\,\,\,\,\,\,\, 10^{-4}\text{Hz}\lesssim f_{7} \lesssim \textcolor{black}{ 10^{-3}}\text{Hz},\\
      &h_{8}\approx \textcolor{black}{4}\times 10^{-21}, \,\,\,\,\,\,\,\,\,\,\,\,\,\,\,\,\,\,\,\,\,\,\,\,\,\,\,\,\,\,\,\,\,\,\,\,\,\,\,\,\,\,\,\,\,\,\, f_{8}\approx 10^{-4}\text{Hz} \,\,\nonumber.
\end{align}
Note that, while the expected strain is similar to that of WDs, the typical frequency in this case is much lower, due to larger BH masses.

\subsection{Gravitational wave background derivation}
The goal of this derivation is to find an expression for the characteristic amplitude $h_{\rm c}$ of the background signal in terms of frequency. In order to do so, following the steps illustrated by \citet{Phinney:01aa} and \citet{Sesana:08aa}, the starting point is the definition of the gravitational energy flux from a distant source, $S(t)$, written as
\begin{align}
    S(t)=\frac{c^3}{16\pi G}\left(\dot{h}^2_{+}+\dot{h}^2_{\times} \right),
    \label{eq:energyflux}
\end{align}
where $c$ is the speed of light, $G$ is the gravitational constant and $h_{+,\times}$ are the two GW polarizations\footnote{We are assuming that the Transverse Traceless gauge holds.}. The dot indicates the time derivative. If we consider the Fourier Transform (FT) of the waveforms
\begin{align}
    \tilde{h}_{+,\times}(f)=\int_{-\infty}^{+\infty}h_{+,\times}\exp{(-i2\pi ft)} dt,
\end{align}
and Parseval's theorem
\begin{align}
    \int_{-\infty}^{+\infty}|h_{+,\times}(t)|^2dt=\int_{-\infty}^{+\infty}|\tilde{h}_{+,\times}(f)|^2df,
\end{align}
we can write the time integral of equation~\eqref{eq:energyflux} as
\begin{align}
    \int_{-\infty}^{+\infty}dt S(t)=\frac{c^3\pi}{2G}\int_{0}^{+\infty} df f^2 \left(|\tilde{h}_{+}(f)|^2+|\tilde{h}_{\times}(f)|^2 \right),
\end{align}
where the integration domain has changed from $(-\infty,+\infty)$ to $[0,+\infty)$ thanks to the symmetry properties of the FT. If we take the average of the energy flux over all the possible orientations of the source, $\Omega_{\rm s}$, we get
\begin{align}
    <S(t)>_{\Omega_{\rm s}}=\frac{L_{\rm GW}(t)}{4\pi d^2_{ \rm L}},
    \label{eq:orientationflux}
\end{align}
where $d_{\rm L}$ is the luminosity distance and $L_{\rm GW}$ is the emitted GW luminosity measured in the rest frame of the source. The time integral of equation~\eqref{eq:orientationflux} is simply
\begin{align}
    \int_{-\infty}^{+\infty}dt <S(t)>_{\Omega_{\rm s}}=\frac{(1+z)}{4\pi d^2_{\rm L}}\int_{-\infty}^{+\infty}L_{\rm GW}(t_{\rm r})dt_{\rm r}=\frac{(1+z)}{4\pi d^2_{\rm L}}E_{\rm GW},
    \label{eq:integralflux}
\end{align}
where $E_{\rm GW}$ is the rest-frame GW energy, $z$ is the redshift and $t_{\rm r}$ is the time local to the event, related to the observed time, $t$, by 
\begin{align}
    t=(1+z)t_{\rm r}.
    \label{eq:resttime}
\end{align}
From the above calculations, we derive 
\begin{align}
    &\int_{-\infty}^{+\infty}dt <S(t)>_{\Omega_{\rm s}}= \nonumber\\
    &=\frac{c^3\pi}{2G(1+z)}\int_{0}^{+\infty}df_{\rm r}f^2<\left(|\tilde{h}_{+}(f)|^2+|\tilde{h}_{\times}(f)|^2 \right)>_{\Omega_{\rm s}}\nonumber\\
    &=\frac{1+z}{4\pi d^2_{\rm L}}\int_{0}^{+\infty}\frac{dE_{\rm GW}}{df_{\rm r}}df_{\rm r},
    \label{eq:denergy}
\end{align}
where $f_{\rm r}$ is the rest-frame frequency that can be expressed in terms of the observed frequency, $f$, as $f_{\rm r}=(1+z)f$ (this follows immediately from equation~\ref{eq:resttime}). Thus, we see from equation~\eqref{eq:denergy} that the emitted GW energy per bin of rest-frame frequency is
\begin{align}
  \frac{dE_{\rm GW}}{df_{\rm r}}=\frac{2\pi^2c^3\textcolor{black}{d}^2}{G}\textcolor{black}{f^2}<\left(|\tilde{h}_{+}(f)|^2+|\tilde{h}_{\times}(f)|^2 \right)>_{\Omega_{\rm S}},
  \label{enq:energyperfreq}
\end{align}
\textcolor{black}{where $d$} is related to $d_{\rm L}$ in the following way (if $\Omega_{\rm k}=0$, see \citealt{Hogg:99aa})
\begin{align}
  \textcolor{black}{d}=\frac{d_{\rm L}}{1+z}.
\end{align}
\textcolor{black}{Since we are averaging over the angles and since previous studies (e.g. \citealt{Kobayashi_2004}) have shown that $h_{+}\sim h_{\times}\sim h$, we consider 
\begin{align}
  \tilde{h}_{+}(f)\approx \tilde{h}_{\times}(f)\approx \tilde{h}(f) ,
\end{align}
and so we can write
\begin{align}
  \frac{dE_{\rm GW}}{df_{\rm r}}=\frac{4\pi^2c^3d^2}{G}\textcolor{black}{f^2}|\tilde{h}(f)|^2 . 
   \label{eq:densitywithh}
\end{align}
\indent Until now we have considered the signal from a single source. Since we are interested in the signal from the entire population, we proceed in the following way. We introduce the GW present-day energy density, $\mathcal{E}_{\rm GW}$, given by \citep{Phinney:89aa}
\begin{align}
   \mathcal{E}_{\rm GW}= \int_{0}^{+\infty}S_{\rm E}(f)df,
   \label{eq:firstdensity}
\end{align}
where $S_{\rm E}$ is the spectral energy density of the background \citep{Moore:15aa}
\begin{align}
    S_{\rm E}=\frac{\pi c^2}{4G}fh_{\rm c}^2,
\end{align}
with $h_{\rm c}$ characteristic strain \citep{Maggiore:07aa}
\begin{align}
    |h_{\rm c}(f)|^2=4f^2|\tilde{h}(f)|^2.
\end{align}
Thus, equation \eqref{eq:firstdensity} can be written as
\begin{align}
    \mathcal{E}_{\rm GW}=\int_{0}^{+\infty}\frac{\pi c^2}{4G}f^2h^2_{\rm c}(f)\frac{df}{f}.
    \label{eq:en_dens1}
\end{align}
Assuming that the Universe is isotropic and homogeneous, $\mathcal{E}_{\rm GW}$ is equal to the sum of the energy densities emitted from the single sources at each redshift
\begin{align}
   \mathcal{E}_{\rm GW}= \int_{0}^{+\infty}dz\frac{d\#}{dtdz}\frac{1}{c}\left(\int_{-\infty}^{+\infty}dt<S(t)>_{\Omega_{\rm s}}\right),
   \label{eq:en_dens2}
\end{align}
where $d\#/dtdz$ is the number of sources generating GWs in the observed time $dt$, inside the redshift interval $[z,z+dz]$}. Comparing equations~\eqref{eq:en_dens1} and \eqref{eq:en_dens2}, and using equation \eqref{eq:denergy}, we finally obtain the following formula for the characteristic strain
\begin{align}
    h_{\rm c}^2=\frac{G}{c^3\pi^2}\frac{1}{f}\int_0^{+\infty}dz\frac{d\#}{dtdz}\textcolor{black}{\frac{1}{d^2}}\left(\frac{dE_{\rm GW}}{df_{\rm r}}\right)_{\big\rvert_{f_{\rm r}=f(1+z)}}.
    \label{eq:carstrain}
\end{align}

\subsection{Order of magnitude estimates of the background signal}
Before developing the calculations in a more formal way, we can give an estimate of the GW background from MS stars and WDs in the following way. We approximate the FT of the strain as $h/f$ (we will justify why it is possible to do this in section \ref{sec:3}) and we write
\begin{align}
    \frac{d\#}{dtdz}\approx \frac{dN^{\rm tde}}{dtdz}\approx \frac{\dot{N}^{\rm tde}_{\rm gal}}{1+z}\frac{dN^{\rm gal}}{dz},
\end{align}
where $N^{\rm tde}$ is the number of tidal disruption events, $\dot{N}^{\rm tde}_{\rm gal}$ is the rate of TDEs per galaxy and $N^{\rm gal}$ is the number of galaxies. Thus, we can write equation \eqref{eq:carstrain} as
\begin{align}
    h_{\rm c}^2&\approx 4 \times \frac{1}{f}\times \dot{N}^{\rm tde}_{\rm gal}\times h^2\times \int_0^{+\infty}dz\frac{dN_{\rm gal}}{dz}\frac{1}{1+z}.
    \label{eq:estimate}
\end{align}
If we consider a Sun-like star disrupted by a $10^6\,\text{M}_{\sun}$ BH, we have that the frequency is $\approx 10^{-4}\,\text{Hz}$ (cf. section \ref{par:TDE_theory}) and the TDE rate is $\dot{N}^{\rm tde}_{\rm gal}\approx 10^{-4}\text{yr}^{-1}\,\text{gal}^{-1}$ (see, e.g., \citealt{Stone:16aa}). If we consider about 0.01 galaxies per unit of cubic Megaparsec (see, e.g, \citealt{Montero-Dorta:09aa}), we can estimate that up to $z\sim 1$ (i.e. $4\times 10^{3}\,\text{Mpc}$) there are
\textcolor{black}{$0.01\,\text{gal}/\text{Mpc}^{3}\times (4\times 10^3\,\text{Mpc})^{3}\approx 10^{8}\,\text{galaxies}$}. 
Inserting all this information in equation \eqref{eq:estimate}, we get
\begin{align}
    h_{\rm c}\approx \sqrt{10}h.
\end{align}
So we expect the GW background from TDEs of MS stars to be around the same order of magnitude of the the strain from the single event. Moreover, if we compare the background from MS stars and the one from WDs we obtain
\begin{align}
    \frac{h^2_{\rm c, MS}}{h^2_{\rm c, WD}}\approx \frac{f_{\rm WD}}{f_{\rm MS}}\times \frac{\dot{N}^{\rm MS}_{\rm gal}}{\dot{N}^{\rm WD}_{\rm gal}}\times\frac{h^2_{\rm MS}}{h^2_{\rm WD}},
\end{align}
 and considering a MS star as in the previous example, and a WD disrupted by a \textcolor{black}{$10^{5}\,\text{M}_{\sun}$} BH with an estimated rate of $\dot{N}^{\rm WD}_{\rm gal}\approx \textcolor{black}{10^{-4}}\,\text{/y/gal}$ (see, e.g., \citealt{Stone:16aa}), we have
\begin{align}
\textcolor{black}{h^2_{\rm c, MS} \approx h^2_{\rm c, WD}}.
\end{align}
So the background of Sun-like stars disrupted by a $10^6\,\text{M}_{\sun}$ BH is around the same order of magnitude as the background of WDs disrupted by a $\textcolor{black}{10^5}\,\text{M}_{\sun}$ BH and, since both the signals are not very strong, we do not expect them to be detected (at least for LISA and TianQin). \textcolor{black}{However, if we assume that these WDs are disrupted not in galactic nuclei but by IMBHs residing in globular clusters, with a disruption rate around $\dot{N}^{\rm WD}_{\rm gc}\approx \textcolor{black}{10^{-3}}/\text{y}/\text{gc}$ (see, e.g., \citealt{Baumgardt:04aa}), we get}
\begin{align}
\frac{h_{\rm c, MS}}{h_{\rm c, WD}}\approx \left(\frac{f_{\rm WD}}{f_{\rm MS}}\times \frac{\dot{N}^{\rm MS}_{\rm gal}}{\dot{N}^{\rm WD}_{\rm gc}}\times\frac{h^2_{\rm MS}}{h^2_{\rm WD}}\times\frac{1}{N_{\rm gc}}\right)^{1/2}\approx \textcolor{black}{\left(10N_{\rm gc}\right)}^{-1/2}.
\end{align}
Thus, if we take into account WDs disrupted by IMBHs in globular clusters, the estimated number of GCs per galaxy becomes a key factor in the derivation of this background. \textcolor{black}{Now that we have explored the expected magnitude of the background, we move to a full description of the physical scenario at hand.}

\section{Methods}
To derive the GW background signal we need to specialize two terms in equation \eqref{eq:carstrain}: the GW energy per unit frequency and the number of sources per unit time per unit redshift.
\label{sec:3}

\subsection{Number of sources per unit time per unit redshift: white dwarfs}
We need to find the proper expression for $d\#/dtdz$. Since we assume a fixed stellar mass, the other possible variables that this quantity can depend on, apart from $z$ and $t$, are the mass of the IMBH in the GC, $M_{\rm h}$, and the number of GCs per galaxy, $N_{\rm gal}^{\rm gc}$. We assume that the mass distribution of IMBHs is a $\delta$ function at a fixed value of $M_{\rm h}$, that we take as a free parameter in the range $10^3 \text{M}_{\sun}\leq M_{\rm h} \leq 10^{5}\text{M}_{\sun}$. So, the only variable left is $N_{\rm gal}^{\rm gc}$. It is reasonable to assume that there is a scale relation between this quantity and the luminosity of the galaxy that hosts the GCs. Since this luminosity can be more conveniently expressed in terms of the mass of the SMBH in the nucleus of the galaxy, $M_{\bullet}$ (see, e.g., \citealt{Faber:76aa} and \citealt{Ferrarese:00aa}), we can write
\begin{align}
    \frac{d\#}{dtdz}\longrightarrow\int dM_{\bullet}\frac{d\#}{dtdzdM_{\rm \bullet}}.
    \label{eq:totrate}
\end{align}
In particular, equation~\eqref{eq:totrate} can be expressed as
\begin{align}
    \int dM_{\bullet}\frac{d\#}{dtdzdM_{\rm \bullet}}=\dot{N}_{\rm gc}^{\rm tde}\int dM_{\bullet} N_{\rm gal}^{\rm gc}\frac{dn}{dM_{\rm \bullet}}\frac{1}{1+z}\frac{dV_{\rm c}}{dz},
    \label{eq:totratefull}
\end{align}
where $\dot{N}_{\rm gc}^{\rm tde}$ is the rate of TDEs of WDs per globular cluster, $N_{\rm gal}^{\rm gc}$ is the number of globular clusters per galaxy, ${dn}/{dM_{\rm \bullet}}$ is the number density of \textcolor{black}{galaxies} per unit SMBH mass and ${dV_{\rm c}}/{dz}$ is the comoving volume per redshift slice $dz$. In the following paragraphs we explicit each term of equation~\eqref{eq:totratefull}.

\subsubsection{Rate of TDEs per GC}
\label{sec:rate}
We can compute the rate of TDEs in globular clusters following the loss cone theory \citep{Frank:76aa}. In particular, as done by \citet{Baumgardt:04aa} and \citet{Baumgardt:04ab} (from now on we will refer to as B04a and B04b), we consider globular clusters where the critical radius $r_{\rm cr}$, i.e. the distance from the black hole where there is the transition from the \textit{full loss cone} to the \textit{empty loss cone} regime, is lower than the influence radius of the hole $r_{\rm i}$. With this assumption, we derive the TDE rate as the ratio between the number of stars in the loss cone at $r_{\rm cr}$ and the crossing time $T_{\rm c}=\textcolor{black}{r/\sigma}$ (with $\sigma$ stellar velocity dispersion, see \citealt{Amaro-Seoane:01aa}) at $r_{\rm cr}$
\begin{align}
    \dot{N}^{\rm tde}_{gc}(r_{\rm cr})=\frac{n(r)r^3\theta^2_{\rm lc}}{T_{\rm c}}\bigg\rvert_{r_{\rm cr}},
\end{align}
where we have introduced the stellar number density, $n(r)$, and the opening angle of the loss cone, $\theta_{\rm lc}$ given by \citep{Frank:76aa}
\begin{align}
    \theta_{\rm lc}=f\left(\frac{2r_{\rm t}}{3r}\right)^{1/2},
\end{align}
\textcolor{black}{with $f\approx 2$}. In general $n(r)$ is written as
\begin{equation}
    n(r)=n_{0}r^{-\textcolor{black}{\eta}},
\end{equation}
where $n_0$ is the cusp density and $\textcolor{black}{\eta}$ is the power law index. We take $\textcolor{black}{\eta}=1.75$, which is the value used by B04b for compact objects. We use the following relations (see \citealt{Frank:76aa} and B04a) for the critical radius
\begin{align}
    r_{\rm cr}=0.2\left(\frac{r_{\rm t}M_{\rm h}^{2}}{M^2_* n_0} \right)^{4/9}, 
\end{align}
and the influence radius
\begin{align}
  r_{\rm i}=\frac{3}{8\pi} \frac{M_{\rm h}}{M_*n_{\rm c}r^2_{\rm c}},
\end{align}
where we have introduced the core density $n_{\rm c}$ and the core radius $r_{\rm c}$, assuming that the cusp density flattens into a constant core density at $r=2r_{\rm i}$ (see B04a). Putting all together, we obtain the following expression for the rate\footnote{\textcolor{black}{B04a and B04b multiply their theoretical rate by a constant $k_{\rm D}$ that they get doing a best fit from the data of their simulations. We do not take into account this constant.}}
\begin{align}
    \dot{N}^{\rm tde}_{\rm gc}\sim \textcolor{black}{60}\text{Myr}^{-1}&\left(\frac{R_{\rm WD}}{\text{R}_{\sun}}\right)^{4/9}\left( \frac{M_{\rm WD}}{\text{M}_{\sun}}\right)^{-95/54}\times\nonumber\\
    &\left(\frac{M_{\rm h}}{10^3\text{M}_{\sun}}\right)^{61/27}\left( \frac{n_{\rm c}}{\text{pc}^{-3}}\right)^{-7/6}\left(\frac{r_{\rm c}}{\text{1pc}} \right)^{-49/9}.
\end{align}
Since this rate is derived at the critical radius, we can use it as a good estimate for both the full loss cone and the empty loss cone regime. But there is an important difference between these two scenarios. While in the empty loss cone regime the stars are typically disrupted at the tidal radius (i.e. $\beta \approx 1$), in the full loss cone regime stars can cross the loss cone many times before being completely disrupted, allowing for a larger range of $\beta$ factors. In this latter regime, we therefore need to consider the distribution of $\beta$ factors in the derivation of the TDE rate. In particular, we assume that the rate of TDEs can be written as
\begin{align}
    \frac{d\dot N}{d\beta}=\frac{\dot{N}^{\rm tde}_{\rm gc}}{\beta^{\textcolor{black}{\gamma}}},
    \label{eq:betadistr}
\end{align}
\textcolor{black}{and we take $\gamma=2$} (following \citealt{Stone:16aa}) from which it follows 
    \begin{align}
   \dot{N}^{\rm tde}_{\rm gc} =\int_{1}^{+\infty}d\beta \beta^{-2} \dot{N}^{\rm tde}_{\rm gc},
\end{align}
that holds since the distribution of $\beta$ is normalized to 1. \textcolor{black}{In the calculations we take $r_{\rm t}/r_{\rm s}$ as upper limit for $\beta$.} 
\subsubsection{Number of GCs per galaxy}
\citet{Harris:11aa}, following-up of the work of \citet{Burkert:10aa}, suggest this scaling relation between the number of GCs and the SMBH mass in a galaxy
\begin{align}
    N^{\rm gc}_{\rm gal}= 10^{(-5.78\pm 0.85)} \left( \frac{M_{\bullet}}{M_{\sun}}\right)^{(1.02\pm 0.10)}.
    \label{eq:numbergcs}
\end{align}
Forcing the slope of the line to be 1, they get the following best fit relation
\begin{align}
     N^{\rm gc}_{\rm gal}=N_{0} \left( \frac{M_{\bullet}}{4.07\times 10^5 M_{\sun}}\right)^{\lambda},
     \label{eq:ngc}
\end{align}
where they have both the parameters $N_0$ and $\lambda$ equal to 1. These relations are obtained from a study on a sample of 33 galaxies and in particular they find that this scaling is appropriate for elliptical and spiral galaxies, but not for lenticular ones, that seem not to follow a particular trend. Still, they discover that $10\%$ of the galaxies in their sample strongly deviate from this relation, in the sense that their SMBH mass is ten time smaller than the one predicted by this relation and this deviation cannot be solved within the uncertainties. Between these \textit{problematic} galaxies there is also the Milky Way (MW). We know that the MW has around $\sim 160$ GCs and the mass of its SMBH, Sgr A*, is around $\sim 4\times 10^6\text{M}_{\sun}$. To adjust a scaling such that of equation \eqref{eq:ngc} to match the MW properties, we can proceed in the following ways: (i) we can take $\lambda=2.2$ and $N_0=1$
\begin{align}
   N^{\rm gc}_{\rm gal} = \left( \frac{4\times 10^6 \text{M}_{\sun}}{4.07\times 10^5 M_{\sun}}\right)^{2.2}\approx 160,
\end{align}
or (ii) we can take $\lambda=1$ and $N_0=16$ 
\begin{align}
    N^{\rm gc}_{\rm gal}= 16 \left( \frac{4\times 10^6 \text{M}_{\sun}}{4.07\times 10^5 M_{\sun} }\right) \approx 160.
\end{align}
Thus, for the derivation of the background we consider $\lambda$ and $N_{0}$ as free parameters that can change in the following intervals: $\lambda \in [1,2.2]$ and $N_{0} \in [1, 16]$. 
\subsubsection{Galaxy distribution}
\label{sec:luminosity}
To derive $dn/dM_{\bullet}$, we start from the Schechter luminosity function \citep{Schechter:75aa} in the R-band
\begin{align}
    \frac{dn}{dL_{\rm R}}=\frac{\phi_*}{\textcolor{black}{L_{\rm *}}}\left(\frac{L_{\rm R}}{L_*}\right)^{\alpha}\exp{\left(-\frac{L_{\rm R}}{L_*}\right)},
\end{align}
where $\phi_*$, $L_*$ and $\alpha$ are quantities that depend on $z$. We parametrize them as done by \citet{Gabasch:06aa} that, for $0 \lesssim z \lesssim 3$, as derived from the Fors Deep Field using the observations collected with the Very Large Telescope (VLT), obtain
\begin{align}
    \alpha& \approx\textcolor{black}{-1.33},\label{eq:alpha}\\
    \phi_*&\approx 0.0037\,\text{Mpc}^{-3}(1+z)^{-0.68},\label{eq:phi*}\\
    L_*&\approx 5\times 10^{10}\text{L}_{\sun}(1+z)^{0.64}.\label{eq:l*}
\end{align}
Using the Faber-Jackson law in the R-band \citep{Faber:76aa}
\begin{align}
  \sigma\approx 150\, \text{km/s}\, \left(\frac{L_{\rm R}}{10^{10}\text{L}_{\sun}}\right)^{1/4},
\end{align}
where $\sigma$ is the central stellar velocity dispersion of the (elliptical) galaxy, and the $M_{\bullet}-\sigma$ relation \citep{Ferrarese:00aa} with the calibrations of \citet{McConnell:13aa}
\begin{align}
    M_{\bullet}=10^{8.32}\text{M}_{\sun}\left(\frac{\sigma}{200\,\text{kms}^{-1}} \right)^{5.64},
\end{align}
we can write the distribution of galaxies in terms of $M_{\bullet}$ as
\begin{align}
    \frac{dn}{dM_{\bullet}}=&\frac{0.0028\,\text{Mpc}^{-3}}{10^8\text{M}_{\sun}(1+z)^{0.48}}\left( \frac{M_{\bullet}}{10^{8}\text{M}_{\sun}}\right)^{-1.17}\times \nonumber\\
    &\exp{\left(-\frac{0.4}{(1+z)^{0.7}}\left(\frac{M_{\bullet}}{10^{8}M_{\sun}}\right)^{0.709}\right)}.
\end{align}
\subsubsection{Comoving volume per unit of redshift}
This quantity is given by (see \citealt{Hogg:99aa})
\begin{align}
\frac{dV_{\rm c}}{dz}=4\pi\frac{c}{H_0}d^2\frac{1}{E(z)},
\end{align}
where $E(z)$ is the dimensionless parameter
\begin{align}
    E(z)=\sqrt{\Omega_{\rm M}(1+z)^3+\Omega_{\Lambda}},
\end{align}
\textcolor{black}{with} $\Omega_{\rm M}$ and $\Omega_{\Lambda}$ dimensionless density parameters for matter and dark energy respectively (assuming $\Omega_{\rm k}=0$). In this work, following \citet{Gabasch:06aa}, we take $\Omega_{\rm M}=0.3$, $\Omega_{\Lambda}=0.7$ and $H_{0}=70 \,\text{Km/(sMpc)}$.
\subsection{Number of sources per unit time per unit redshift: main sequence stars}
For MS stars, the number of sources per unit of time per unit of redshift will depend on the mass of the SMBH involved in the disruption, $M_{\bullet}$, so we can write
\begin{align}
   \frac{d\#}{dtdz}\longrightarrow\int dM_{\bullet}\frac{d\#}{dtdzdM_{\rm \bullet}}=\int d M_{\bullet}\,\dot{N}^{\rm tde}_{\rm gal}\frac{dn}{dM_{\rm \bullet}}\frac{1}{1+z}\frac{dV_{\rm c}}{dz}.
\end{align}
For the term $\dot{N}^{\rm tde}_{\rm gal}$, we need to consider that this rate will depend on both the stellar mass and the distance between the star and the black hole (see section  \ref{sec:rate} for the considerations about full and empty loss cone). If we assume
\begin{align}
    \frac{d\dot N}{d\beta dM_*}=\frac{\dot{N}^{\rm tde}_{\rm gal}}{\beta^2\phi(M_{*})},
\end{align}
where $\phi(M_*)$ is the Salpeter function \citep{Salpeter:55aa} normalized in the interval $[1,+\infty)$
\begin{align}
    \phi(M_*)=1.35\,\text{M}_{\sun}^{1.35}M_*^{-2.35},
\end{align}
we can write
\begin{align}
    \dot{N}^{\rm tde}_{\rm gal}(M_{\bullet})=\int_{1}^{+\infty} dM_* \int_{1}^{+\infty}d\beta \,\beta^{-2}\phi(M_*)\dot{N}^{\rm tde}_{\rm gal}(M_{\bullet})\,.
\end{align}
In the calculations we take $M_*=100\,\text{M}_{\sun}$ and $\beta=r_{\rm t}/r_{\rm s}$ as upper limits for $M_*$ and $\beta$ rispectively. For the rate, we take the one calculated by \citet{Stone:16aa}
\begin{align}
    \dot{N}^{\rm tde}_{\rm gal}=2.9\times 10^{-5}\,\text{yr}^{-1}\left(\frac{M_{\bullet}}{10^8\,\text{M}_{\sun}} \right)^{-0.404}.
\end{align}
\subsection{Gravitational energy per bin of rest frame frequency}
We need to specialize equation \eqref{eq:densitywithh} for TDEs, which means that we have to find the expression for \textcolor{black}{$\tilde{h}(f)$}. In paragraph \ref{par:TDE_theory} we have \textcolor{black}{explained} that it is suitable to approximate this GW signal from TDEs with a burst. Thus, we are in practice considering the strain like a constant function over the interval [-$\tau$/2,$\tau$/2] (where $\tau$ is the duration of the signal), and zero outside. For these reasons, we can write the FT of the signal as
\begin{align}
    \tilde{h}(f) \approx h\int_{-\tau/2}^{\tau/2}\exp{(-i2\pi ft)}dt\approx h\tau\frac{sin(\pi f\tau)}{\pi f\tau}\approx h\tau\approx \frac{h}{\textcolor{black}{f}},
    \label{eq:ft}
\end{align}
so that equation~\eqref{enq:energyperfreq} can be expressed as
\begin{align}
    \frac{dE_{\rm GW}}{df_{\rm r}}=\frac{4\pi^2 c^3}{G}\textcolor{black}{d^2}h^2,
    \label{eq:energy_strain}
\end{align}
where we have considered that the strain does not depend on the orientation of the source for our problem. \textcolor{black}{ Thus, equation \eqref{eq:energy_strain} becomes} 
\textcolor{black}{\begin{align}
  \frac{dE_{\rm GW}}{df_{\rm r}}=\frac{4\pi^2 c^3}{G}\left(\frac{\beta r_{\rm g}r_{\rm s*}}{r_{\rm t}}\right)^2 \propto \textcolor{black}{f_{\rm r}^{4/3}}
  \label{eq:eq:energy_beta}
\end{align}}
\textcolor{black}{(in the last step, we consider $\beta\propto f^{2/3}$, cf. section \ref{sec:2}).}

\section{Results}
\label{sec:4}
\subsection{Background for white dwarfs}
Considering all the steps illustrated in the previous sections, the final formula for the background is
\begin{align}
    h_{\rm c}=\mathcal{A}_{\rm WD}\left(\frac{f}{10^{-2}\,\text{Hz}} \right)^{-1/2},
\end{align}
where $\mathcal{A}_{\rm WD}$ is a model dependent constant given by
\begin{align}
   \mathcal{A}_{\rm WD}^2\approx& 10^{-\textcolor{black}{53}} \times N_0 \times \left(250 \right)^{\lambda}\times \left(\frac{M_{\rm h}}{10^3\text{M}_{\sun}}\right)^{3.59}\times [\beta_{\rm max}-1]\times\nonumber\\
   &\times\int_{0}^{+\infty}\mathcal{I}(z)dz,
    \label{eq:hc}
\end{align}
with 
\begin{align}
\mathcal{I}(z)=\int_{10^{-2}}^{1}d\mathcal{M}\,\mathcal{M}^{-1.17+\lambda}\exp{\left(-\frac{0.4}{(1+z)^{0.7}}\mathcal{M}^{0.709}\right)}\frac{(1+z)^{\textcolor{black}{-1.48}}}{E(z)}.
\label{eq:ized}
\end{align}
In the calculation, we have taken a GC core density $n_{\rm c}\approx 10^{5}\,\text{pc}^{-3}$ and core radius $r_{\rm c}\approx 0.5\,\text{pc}$. The \textit{mass} dimensionless variable $\mathcal{M}=M_{\bullet}/10^{8}\text{M}_{\sun}$ ranges in the interval $10^{-2}\leq \mathcal{M}\leq 1$, spanning the mass range $10^{6}\text{M}_{\sun}-10^{
8}\text{M}_{\sun}$. The term in square brackets results from the integral over the parameter $\beta$, so we have this term only when considering the full loss cone scenario. \textcolor{black}{We report in Tables \ref{tab:empty_losscone}-\ref{tab:full_losscone} the typical values of $\mathcal{A}_{\rm WD}$ for different choices of the parameters, in the empty and full loss cone, respectively.}\\
\indent Let us estimate the frequency range we expect the signal to cover. The observed frequency is related to the rest frame frequency by $f=f_{\rm r}/(1+z)$. So, when the redshift is zero, the observed frequency coincides with the frequency of the source, meanwhile for higher redshifts $f$ decreases, approaching 0 for $z\rightarrow \infty$. 
\begin{figure}
    \centering
    \includegraphics[width=\columnwidth]{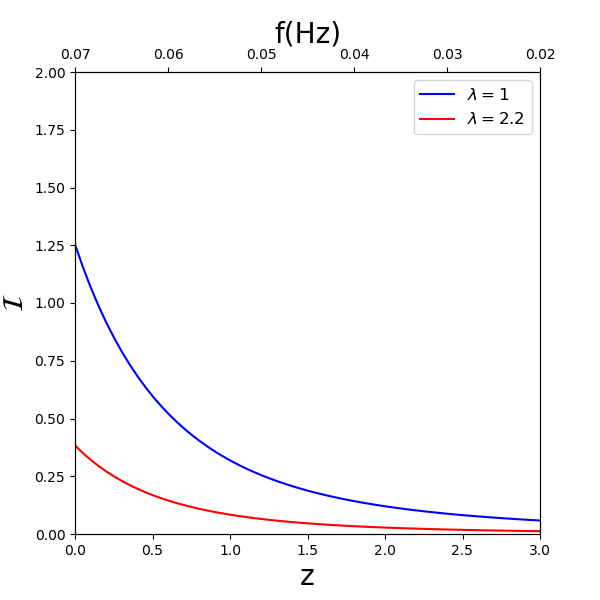}
    \caption{$\mathcal{I}(z)$, i.e. the function under integral \eqref{eq:hc} integrated only in $\mathcal{M}$, plotted with respect to $z$ and $f$. We see that, independently from the value of $\lambda$, this function goes to zero for $z_{\rm max}\approx 3$, i.e. $f\approx 0.02\,\text{Hz}$.}
    \label{fig:1}
\end{figure}
However, although in the above integral we consider $0\leq z < \infty$, from a physical point of view we expect most of the signal to be collected from within a finite redshift $z_{\rm max}$ that we derive by 
inspecting $\mathcal{I}(z)$ in equation \eqref{eq:ized}. This function is plotted in Figure \ref{fig:1} for two extreme values of the parameter $\lambda$, i.e. $\lambda=1$ and $\lambda=2.2$, setting $f_{\rm r}= 0.07\,\text{Hz}$.
\begin{figure}
    \centering
    \includegraphics[width=\columnwidth,, height=0.26\paperheight]{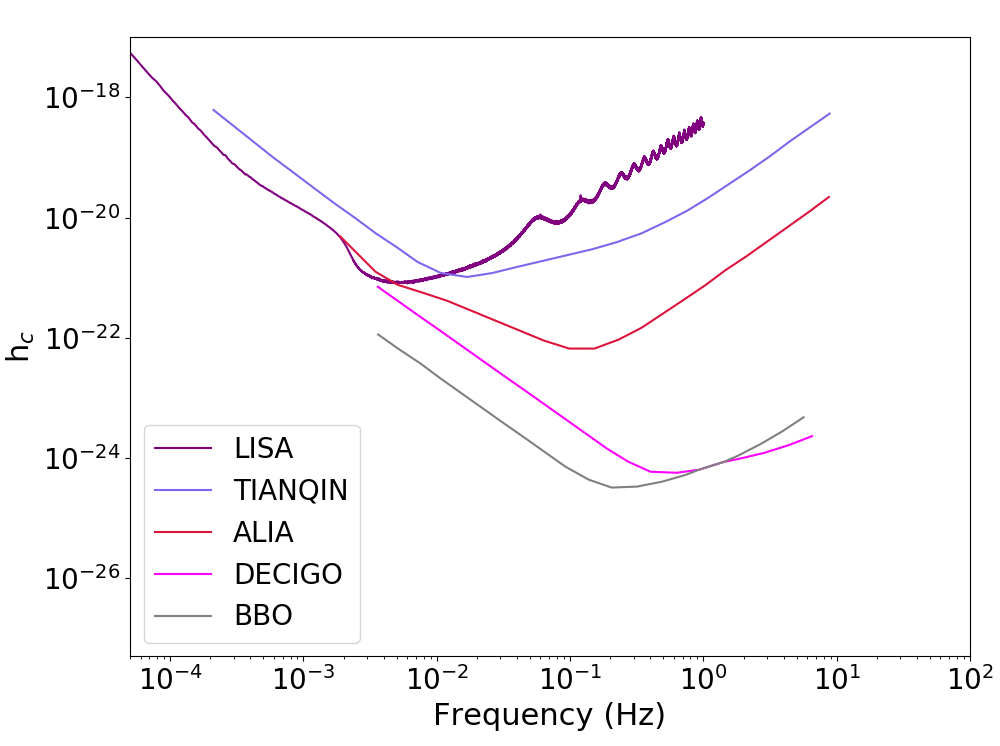}
    \caption{Sensitivity curves of LISA (purple), TianQin (light violet), ALIA (red), DECIGO (pink) and BBO (grey).}
    \label{fig:2}
\end{figure} 
The function vanishes for $z \ge 3$, thus the lowest frequency at which we observe the signal is $f\approx 0.07/4 = 0.02 ~\text{Hz}$. We therefore expect the  frequency interval of the background signal in the empty loss cone regime to be $[0.02 ~\text{Hz},0.07 ~\text{Hz}]$. In the full loss cone scenario $\beta$ is not fixed to 1, so in order to derive the observed frequency interval, we need to consider both $z_{\rm max}$ and the maximum value of $\beta$ (see paragraph \ref{sec:2}). The result is the following
\begin{align}
    &f\in[0.02\text{Hz},11\text{Hz}]\,\,\,\,\,\,\text{for}\,M_{\rm h}=10^3\text{M}_{\sun},\\
    &f\in[0.02\text{Hz},1\text{Hz}] \,\,\,\,\,\,\,\,\,\text{for}\,M_{\rm h}=10^4\text{M}_{\sun},\\
    &f\in[0.02\text{Hz},\textcolor{black}{0.1}\text{Hz}] \,\,\,\,\,\text{for}\,M_{\rm h}=10^5\text{M}_{\sun}.
\end{align}
\begin{figure*}
    \centering
    \includegraphics[width=0.48\textwidth]{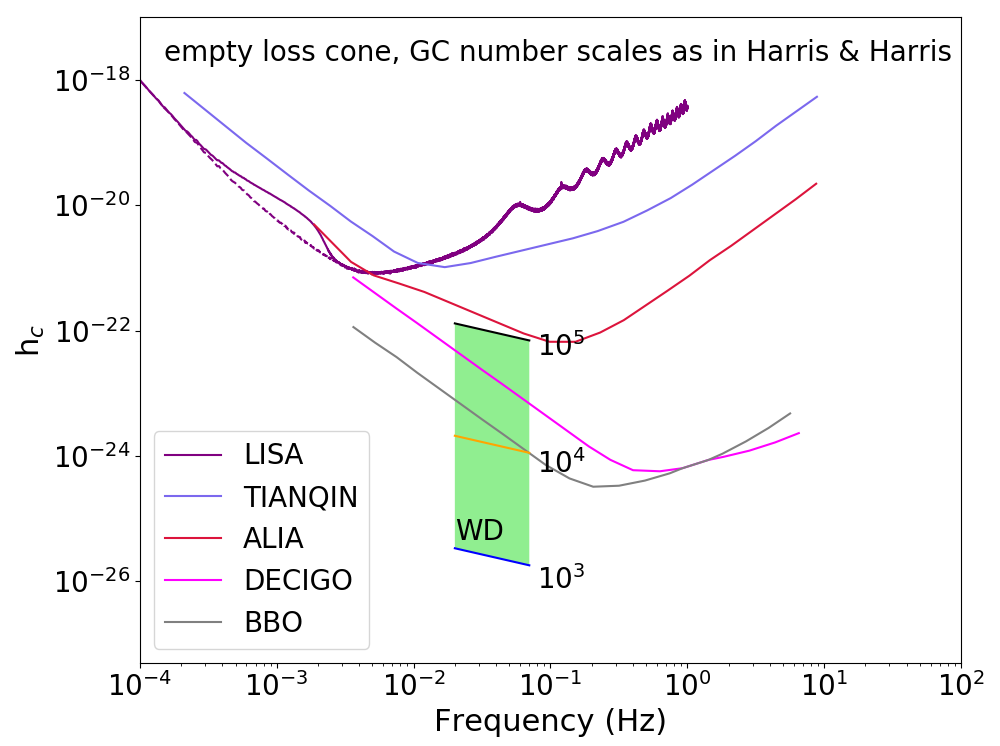}
\includegraphics[width=0.49\textwidth]{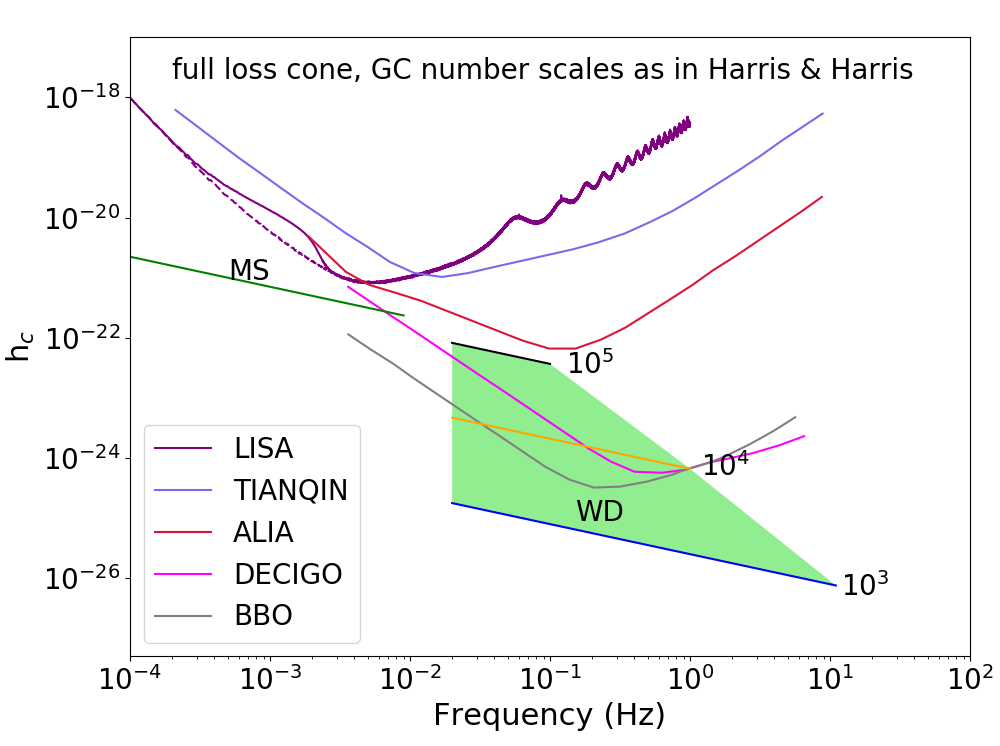}
\caption{GW background, $h_{\rm c}$, plotted with respect to the observed frequency, $f$. We consider the GC number that scales linearly with the mass of the \textcolor{black}{SMBHs} as in \citet{Burkert:10aa} and in \citet{Harris:11aa} (i.e. $\lambda=1$ and $N_{0}=1$). On the left side we show the empty loss cone scenario, on the right side the full loss cone scenario. The blue, orange and black lines represent the background from WDs tidally disrupted by IMBHs of mass $M_{\rm h}=10^3,10^4,10^5\,\text{M}_{\sun}$ respectively. The green area between them stands for all the IMBH masses in between these values. The background is compared with the sensitivity curves of LISA, TianQin, ALIA, DECIGO and BBO. The dark green solid line in the plot on the right side is the background from MS stars.} 
    \label{fig:fig3}
\end{figure*}
Note that for $M_{\rm h}=10^5\text{M}_{\sun}$, the range of frequency is almost the same as for the empty loss cone since the allowed $\beta$ range is small and around $\beta \approx 1$. Indeed, $M_{\rm h}=10^5\text{M}_{\sun}$ is a limit situation for the full loss cone scenario, but we have decided to include also this case for completeness. It follows that the natural frequency range for this physical system goes from the decihertz to a few hertz. For this reason, we compare our results with the sensitivity curves of the following (future) space interferometers: LISA \citep{Amaro-Seoane:17aa}, TianQin \citep{Luo:16aa}, ALIA (\citealt{Bender:13aa} and \citealt{Baker:19aa}), DECIGO \citep{Sato:17aa} and the BBO \citep{Harry:06aa}, that are all shown in Figure \ref{fig:2}.\\
\indent Our results are illustrated in Figures \ref{fig:fig3}-\ref{fig:fig4}-\ref{fig:fig5}. We explore how the background varies if we change the parameters $M_{\rm h}, \lambda\,\,\text{and}\,\,N_{0}$. In Figure \ref{fig:fig3} we illustrate the results for the (pessimistic) case $\lambda=1$ and $N_{0}=1$, that is the scenario where we assume the same linear relation between the number of globular clusters in a galaxy and the mass of the super massive black hole in its core as in \citet{Burkert:10aa} and in \citet{Harris:11aa}. Then, we show the results also for the more optimistic case where $\lambda$ and $N_{0}$ satisfying the conditions for the MW, in particular $\lambda=2.2$ and $N_{0}=1$ in Figure \ref{fig:fig4}, while $\lambda=1$ and $N_{0}=16$ in Figure \ref{fig:fig5}, since these are the two cases where, for a fixed $M_{\bullet}$, we have the highest number of GCs. In each figure, in the left plot we show the empty loss cone scenario, while the full loss cone is illustrated in the right plot. The blue, orange and black \textcolor{black}{lines} represent the GW background if we assume that all the IMBHs have same mass equal to $10^3,10^4,10^5\,\text{M}_{\sun}$ respectively. The green area represents the cases with a IMBH mass in between these values. So the \textit{actual} background of these events will be inside this green area. In all the cases illustrated, the GW background may be detected by DECIGO and it is always above BBO sensitivity curve. \textcolor{black}{It may also be visible to ALIA for the cases $\lambda=2.2,\,N_0=1$ and $\lambda=1, \,N_0=16$, if we consider high BH masses. Moreover, in the most optimistic scenario ($\lambda=2.2,\, N_{0}=1$), the signal even grazes TianQin sensitivity curve.} \\
\indent These results suggest that the TDE background signal from WDs could be indeed detected by both DECIGO and BBO, maybe even by ALIA and surely by any interferometers that will work in the decihertz band with a higher sensitivity than the planned ones.
\vspace{0.2 cm}
\begin{figure*}
    \centering
    \includegraphics[width=0.48\textwidth]{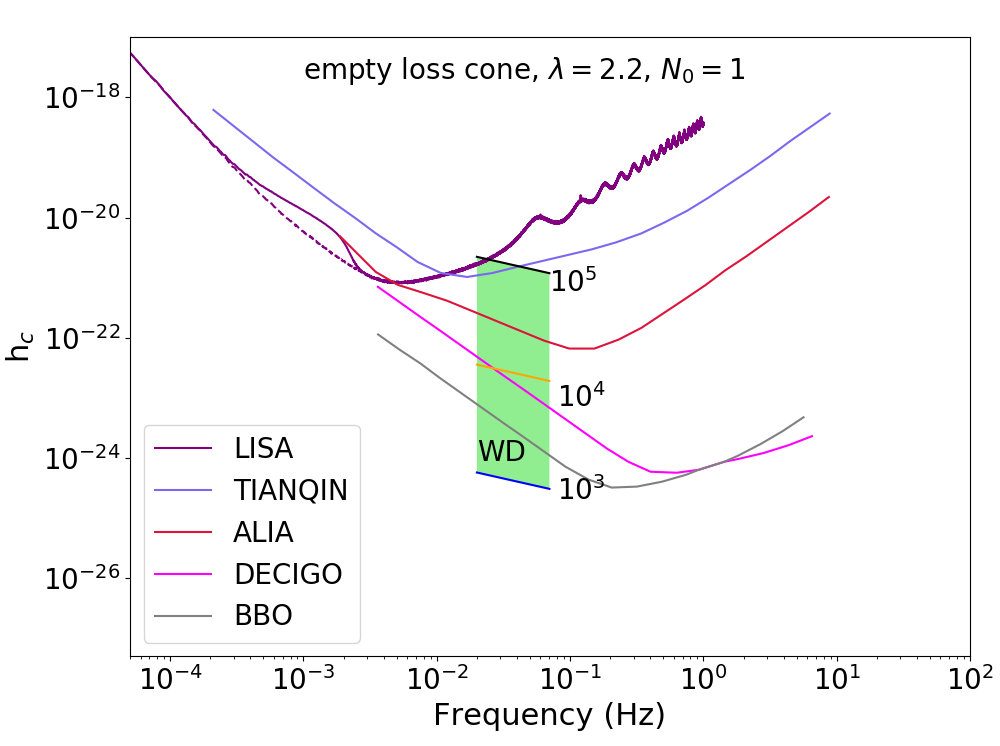}
\includegraphics[width=0.48\textwidth]{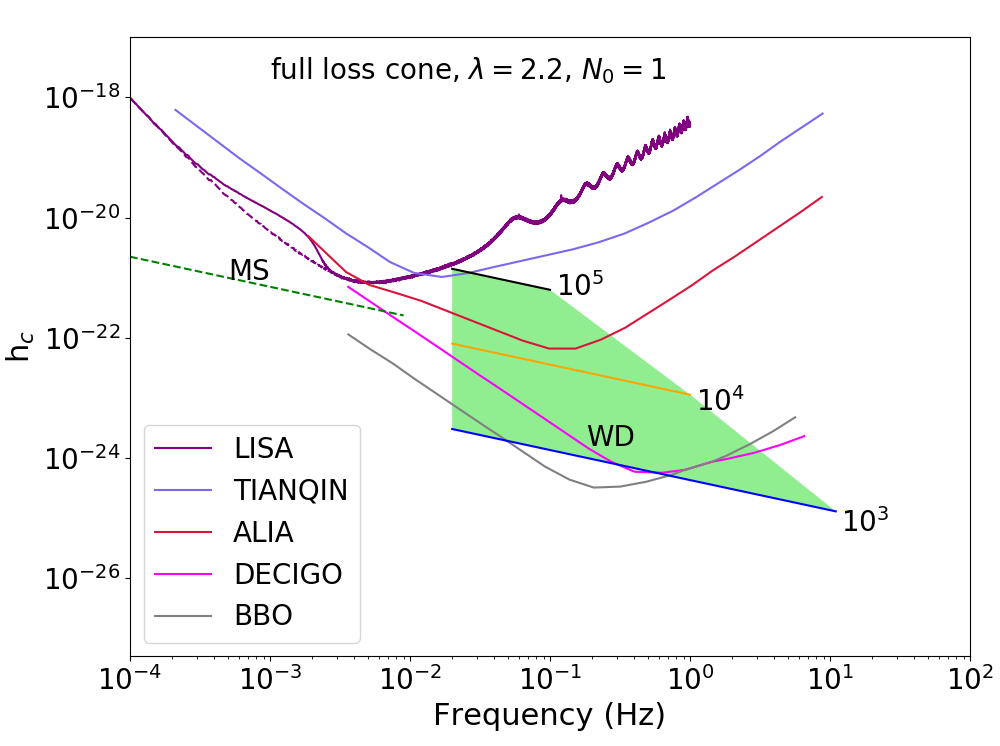}
\caption{GW background, $h_{\rm c}$, plotted with respect to the observed frequency, $f$. We consider $\lambda=2.2$ and $N_{0}=1$. On the left side we show the empty loss cone scenario, on the right side the full loss cone scenario. The blue, orange and black lines represent the background from WDs tidally disrupted by IMBHs of mass $M_{\rm h}=10^3,10^4,10^5\,\text{M}_{\sun}$ respectively. The green area between them stands for all the IMBH masses in between these values. The background is compared with the sensitivity curves of LISA, TianQin, ALIA, DECIGO and BBO. The dark green dashed line in the plot on the right side is the background from MS stars.} 
    \label{fig:fig4}
\end{figure*}

\begin{figure*}
    \centering
    \includegraphics[width=0.48\textwidth]{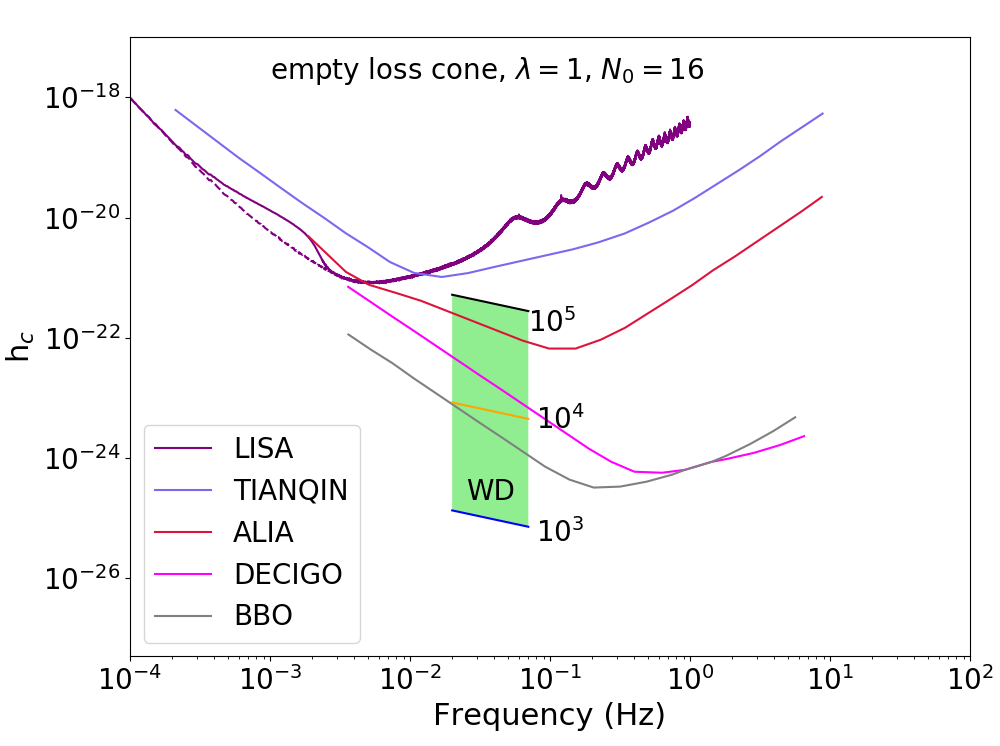}
\includegraphics[width=0.48\textwidth]{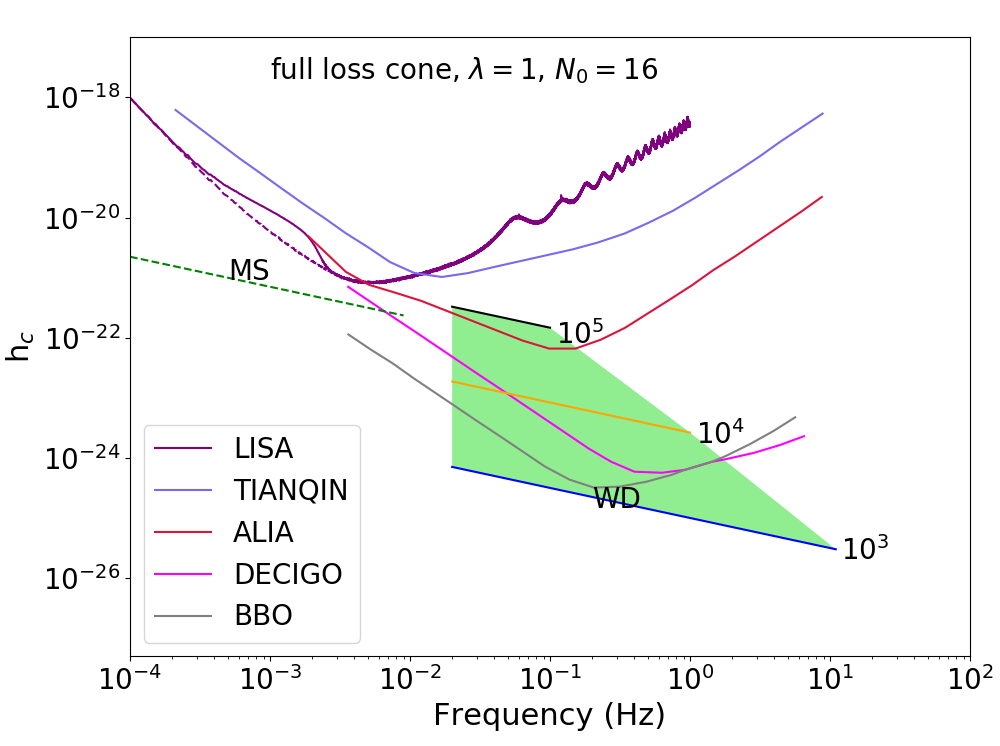}
\caption{GW background, $h_{\rm c}$, plotted with respect to the observed frequency, $f$. We consider $\lambda=1$ and $N_{0}=16$. On the left side we show the empty loss cone scenario, on the right side the full loss cone scenario. The blue, orange and black lines represent the background from WDs tidally disrupted by IMBHs of mass $M_{\rm h}=10^3,10^4,10^5\,\text{M}_{\sun}$ respectively. The green area between them stands for all the IMBH masses in between these values. The background is compared with the sensitivity curves of LISA, TianQin, ALIA, DECIGO and BBO. The dark green dashed line in the plot on the right side is the background from MS stars.} 
    \label{fig:fig5}
\end{figure*}

\begin{table}
	\centering
	\caption{\textcolor{black}{Values of the constant $\mathcal{A}_{\rm WD}$ for different values of $M_{\rm h}$, $\lambda$ and $N_0$. Empty loss cone scenario.}}
	\label{tab:empty_losscone}
	\begin{tabular}{lccr} 
		\hline
		$\text{M}_{\rm h}$&$\mathcal{A}_{\rm WD}$ & $\mathcal{A}_{\rm WD}$ &$ \mathcal{A}_{\rm WD}$ \\
		&$\lambda=1$ $N_0=1$ & $\lambda=2.2$ $N_0=1$ & $\lambda=1$ $N_0=16$\\
		\hline
		$10^3\text{M}_{\sun}$& $5\times 10^{-26}$ & $9\times 10^{-25}$ & $2\times 10^{-25}$ \\
		$10^4\text{M}_{\sun}$& $3\times 10^{-24}$ & $5\times 10^{-23}$ & $1\times10^{-23}$ \\
		$10^5\text{M}_{\sun}$& $2\times 10^{-22}$ & $3\times 10^{-21}$ & $8\times 10^{-22}$ \\
		\hline
	\end{tabular}
\end{table}
\begin{table}
	\centering
	\caption{\textcolor{black}{Values of the constant $\mathcal{A}_{\rm WD}$ for different values of $M_{\rm h}$, $\lambda$ and $N_0$. Full loss cone scenario.}}
	\label{tab:full_losscone}
	\begin{tabular}{lccr} 
		\hline
		$\text{M}_{\rm h}$&$\mathcal{A}_{\rm WD}$ & $\mathcal{A}_{\rm WD}$ &$ \mathcal{A}_{\rm WD}$ \\
		&$\lambda=1$ $N_0=1$ & $\lambda=2.2$ $N_0=1$ & $\lambda=1$ $N_0=16$\\
		\hline
		$10^3\text{M}_{\sun}$& $3\times 10^{-25}$ & $5\times 10^{-24}$ & $1\times 10^{-24}$ \\
		$10^4\text{M}_{\sun}$& $7\times 10^{-24}$ & $1\times 10^{-22}$ & $2\times10^{-23}$ \\
		$10^5\text{M}_{\sun}$& $1\times 10^{-22}$ & $2\times 10^{-21}$ & $5\times 10^{-22}$ \\
		\hline
	\end{tabular}
\end{table}
\subsection{Background for main sequence \textcolor{black}{stars}}
\textcolor{black}{In the case of MS stars, the final formula of the background is
\begin{align}
    h_{\rm c}\approx \mathcal{A}_{\rm MS}\left(\frac{f}{10^{-3}\,\text{Hz}}, \right)^{-1/2}
\end{align}
where $\mathcal{A}_{\rm MS}$ is a constant model dependent
\begin{align}
   \mathcal{A}^2_{\rm MS} \approx 10^{\textcolor{black}{-44}} \,\text{M}_{\sun}^{17/25}  \int_{0}^{+\infty} \mathcal{I_{\rm MS}}(z) dz,  
    \label{eq:ms_back}
\end{align}
with
\begin{align}
\mathcal{I_{\rm MS}}=& \int_{0.01}^{1} d\mathcal{M}\int_{1\text{M}_{\sun}}^{100\text{M}_{\sun}}dM_{*}\mathcal{M}^{-0.24}\exp{\left(-\frac{0.4}{(1+z)^{0.7}}\mathcal{M}^{0.709}\right)}\nonumber\\  &\frac{(1+z)^{\textcolor{black}{-1.48}}}{E(z)}M_*^{-1.68}[\beta_{\rm max}(M_{*}, \mathcal{M})-1].
 \end{align}}
Similarly to the WD case, the term in square brackets derives from the integral over $\beta$, so we have this term only in the full loss cone scenario. Here $\beta_{\rm max}$ is a function of the mass of the SMBH and of the stellar mass, unlike in the previous case where it was a fixed parameter.\\
\indent To determine the frequency range of the signal, we proceed in a similar way as for IMBH-WD background. First, we inspect $\mathcal{I_{\rm MS}}$: it vanishes for $z_{\rm max}\approx 3$. Thus, for the empty loss cone ($\beta\approx 1$) we have that the largest interval possible for the rest frame frequency is $10^{-6}\,\text{Hz}\leq f_{\rm r} \leq 10^{-4}\,\text{Hz}$ (see equation \ref{eq:freqkepl}), which means that the largest window for the observed frequency is $2.5\times 10^{-7}\,\text{Hz}\leq f \leq 10^{-4}\,\text{Hz}$. This frequency band is lower than that covered by any planned detector, while it partially overlaps with that covered by the International Pulsar Timing Array (IPTA). However, the background signal lies orders of magnitude below the IPTA sensitivity curve, therefore overall there are currently no prospects for detection. For the full loss cone scenario, the largest interval for the rest frame frequency is $10^{-6}\,\text{Hz}\leq f_{\rm r}\leq 9\text{mHz}$ and so the observed frequency interval where we investigate the signal becomes $10^{-4}\,\text{Hz}\leq f\leq 9\,\text{mHz}$. 
\textcolor{black}{Thus, gathering all these considerations, the final formula for the background is
\begin{align}
   h_{\rm c}\approx 10^{\textcolor{black}{-21}}\left(\frac{f}{10^{-3}\,\text{Hz}} \right)^{-1/2} 
\end{align}}
We plot this background signal in Figure \ref{fig:fig3} (solid dark green line in the right panel) and then also in Figures \ref{fig:fig4} and \ref{fig:fig5} (dashed green line) as a reference. \textcolor{black}{The background generated by MS stars partly overlaps with the frequency band of BBO, that could detect the high frequency part of this signal. DECIGO may see the highest frequency part of this background too if its sensitivity curve will be at least one order of magnitude more sensitive}.

In Figure \ref{fig:fig3}, note that the background generated by MS stars is comparable with the one produced by WDs (if we consider the most massive IMBHs). In the other two scenarios, where we increase the number of GCs, the WD signal becomes stronger up to one order of magnitude as seen in Figure \ref{fig:fig4}.

\section{Discussion}

\label{sec:5}

\subsection{The spectral shape of the GW background by TDEs}

\textcolor{black}{In the previous sections, we have seen that both for MS stars and WDs, the characteristic strain scales as $h_{\rm c}\propto f^{-1/2}$. This dependence is a consequence of two factors:
(1) the choice $\gamma=2$ in the $\beta$ distribution (eq. (\ref{eq:betadistr})) and (2) the impulsive nature of TDE signals, that leads to $\tilde{h}\sim h/f$ (eq. (\ref{eq:ft})). Thus, the derivative of the GW energy with respect to the rest frame frequency in equation \eqref{eq:eq:energy_beta} is proportional to $f^{4/3}$ (since the only term related to frequency that appears in this equation is $\beta^2$, cf. section \ref{sec:2}).}

\textcolor{black}{The combination of these two assumptions give as a result that the integrand in \eqref{eq:carstrain} is independent on $\beta$, so the characteristic strain $\propto f^{-1/2}$.
While the second assumption is related to the very nature of TDEs, the first one depends on the assumed distribution of pericenter distances, $\mbox{d}\dot{N}/\mbox{d}\beta\propto \beta^{-\gamma}$, with $\gamma=2$. For a generic $\gamma$, the spectral shape is
\begin{align}
    h_{\rm c}\propto f^{(-4\gamma+5)/6}.
\end{align}}

\subsection{Detectability}
We discuss here two assumptions that affect our results on the detectability of the IMBH-WD background signal. First, we assume that all the GCs have an IMBH, i.e. the occupation fraction is 1. Since the background scales with the square root of the occupation fraction, the background signal is lower by a factor of 1.4 (3) if the occupation fraction is $50\%$ ($10\%$). On the other hand, we neglected the black hole spin in the computation of the signal and it has been shown that it may grow by one order of magnitude (see \citealt{Kobayashi_2004}), which implies an increment by a factor $\textcolor{black}{10}$ in our curves.


\subsubsection{Comparison with other sources}
In the same frequency interval investigated in this work, also other sources produce their GW background like galactic WD binaries and low-mass SMBH binaries. The signals from these sources may overlap our signal, but they are in general stronger and detectable by LISA.
The TDE signal will become important when moving to higher sensitivities, and in that case it will be essential to disentangle the various contribution to the background, for example by combining the response obtained with interferometers operating at different wavelengths and with different sensitivities. \textcolor{black}{However, we have noticed a distinctive feature of the TDE background in its characteristic spectral shape $\propto f^{-1/2}$, which is unique given the impulsive nature of such events. This makes the signal from TDEs easily distinguishable from the signals produced by other sources which have a different slope.}

\subsection{Doppler shift}
An interesting phenomenon that we have not taken into account in our calculations is the black hole wandering (see, e.g., \citealt{Lin:80aa}) and how this may affect the GW emission from globular TDEs. We do not expect the BH to emit a significant GW signal due to this movement, but it can cause a Doppler shift in the signal from the tidal disruption. To investigate this, we calculate the shift in frequency as the ratio between the velocity of the hole, $\sigma_{\rm h}$, and the speed of light, 
\begin{align}
   \frac{\Delta f}{f_{\rm r}}=\frac{\sigma_{\rm h}}{c}. 
\end{align}
To derive $\sigma_{\rm h}$, we need first of all to consider that in our scenario $r_{\rm cr}<r_{\rm i}$, so we consider only stars bound to the hole, \textit{i.e.} the stars of the cusp. The number of stars in the cusp, $N_{\rm cusp}$, can be estimated as (see \citealt{Young:77a} and \citealt{Bahcall:77aa}) 
\begin{align}
    N_{\rm cusp}\approx 70 \left(\frac{M_{\rm h}}{10^3\text{M}_{\sun}}\right)^3\left(\frac{10\text{km/s}}{\sigma}\right)^4\left(\frac{0.5\text{pc}}{r_{\rm c}}\right)^{2},
\end{align}
where $\sigma$ is the velocity dispersion of stars with respect to the centre of mass of the system, which means that we have
\begin{align}
    &N_{\rm cusp}\approx 10 \,\,\,\,\,\text{for}\,M_{\rm h}=10^3\text{M}_{\sun},\nonumber\\
    &N_{\rm cusp}\approx 500\,\,\text{for}\,M_{\rm h}=10^4\text{M}_{\sun},\\
    &N_{\rm cusp}\approx 10^5\,\,\text{for}\,M_{\rm h}=10^5\text{M}_{\sun}.\nonumber
\end{align}
Since the mass of the cusp, $M_{\rm cusp}$, is given by $M_{\rm cusp}=M_{*}N_{\rm cusp}$, where the average stellar mass for us is $M_*=0.5\text{M}_{\sun}$, we see that the relation $M_{\rm cusp}\ll M_{\rm h}$ holds. Thus, using the equipartition theorem between the kinetic energy of the hole and the kinetic energy of the stars, we get
\begin{align}
    <\sigma_{\rm h}^2>^{1/2}\sim N^{1/2}_{\rm cusp}\left( \frac{M_*}{M_{\rm h}}\right)<\sigma^2>^{1/2}.
    \label{eq:vh}
\end{align}
If we compute the ratio between the velocity of the hole given in equation \eqref{eq:vh} and $c$, we get a number much smaller than 1 (since we expect $\sigma\approx 10\,\text{km}$, so $\sigma_{\rm h}\approx 10^{-6}c$) and so we can conclude that the Doppler shift of our signal due to the wandering of the IMBH is negligible.

\section{Conclusions}
\label{sec:6}
In this paper we have explored the GW background generated by tidal disruption events. We have focused both on MS stars disrupted by SMBHs and on WDs disrupted by IMBHs residing in globular clusters. Then, we have compared these signals with the sensitivity curves of LISA, TianQin, ALIA, DECIGO and BBO. We have found that the background from MS stars is too low to be detected by these instruments with their current design, \textcolor{black}{apart from BBO that may reveal the high frequency background. This could be detected also by DECIGO if its sensitivity curve was one order of magnitude lower}. The detection of this signal will give us unique information about the population of quiescent SMBHs. By contrast, the GW background from WDs is a promising source for DECIGO and BBO and, in part, for ALIA. The detection of this background will provide important clues on the existence of IMBHs, information on their population, on their occupation fraction in GCs and also on the number of GCs per galaxy.

\section*{Acknowledgements}
This article is based upon work from COST Action CA16104 - Gravitational waves, black holes and fundamental physics, supported by COST (European Cooperation in Science and Technology). MT and GL have received funding from the European Union's Horizon 2020 research and innovation program under the Marie Sk\l{}odowska-Curie grant agreement NO 823823 (RISE DUSTBUSTERS project). \textcolor{black}{MT acknowledges the site \url{http://gwplotter.com} for the sensitivity curves of the instruments TianQin, ALIA, DECIGO and BBO.}



\bibliographystyle{mnras}







\bsp	
\label{lastpage}
\end{document}